\documentclass[hyper]{JHEP} 

\usepackage{epsfig}

\newcommand\fverb{\setbox\pippobox=\hbox\bgroup\verb}

\newcommand\fverbdo{\egroup\medskip\noindent%

            \fbox{\unhbox\pippobox}\ }

\newcommand\fverbit{\egroup\item[\fbox{\unhbox\pippobox}]}

\newbox\pippobox

\title{Carroll Limit of Non-BPS Dp-Brane }
\author{J. Kluso\v{n}\\
Department of
Theoretical Physics and Astrophysics\\
Faculty of Science, Masaryk University\\
Kotl\'{a}\v{r}sk\'{a} 2, 611 37, Brno\\
Czech Republic\\
E-mail: \email{klu@physics.muni.cz}} \preprint{}

 \abstract{We find Carroll non-BPS Dp-brane action by performing Carroll limit of a canonical form of  unstable Dp-brane action. We analyze
 different Carroll limits and discuss  solutions of the equations of motion of Carroll non-BPS Dp-brane at the tachyon vacuum.}

\def\tp{\tilde{p}}

\def\tpi{\tilde{\pi}}

\def\tK{\tilde{K}}

\def\bA{\mathbf{A}}

\def\ttau{\tilde{\tau}}

\def\tF{\tilde{F}}
\def\tT{\tilde{T}}

\def\tlambda{\tilde{\lambda}}

\def\tx{\tilde{x}}

\def\be{\begin{equation}}

\def\ee{\end{equation}}

\def\bea{\begin{eqnarray}}

\def\eea{\end{eqnarray}}

\def\tmH{\tilde{\mH}}

\def\mH{\mathcal{H}}

\def\tA{\tilde{A}}

\newcommand{\tZ}{\tilde{Z}}

\newcommand{\tN}{\tilde{N}}

\newcommand{\mK}{\mathcal{K}}

\def \bA{\mathbf{A}}

\newcommand{\ba}{\mathbf{a}}

\newcommand{\mL}{\mathcal{L}}

\def \tZ{\tilde{Z}}

\def\pb #1{\left\{#1\right\}}

\begin{document}
\section{Introduction and Summary}
The holographic ideas have very broad region of applications. For example, it was shown recently that holography is a very useful tool for the analysis of condensed matter systems, for recent review, see \cite{Hartnoll:2016apf}. It turns out that non-relativistic systems play fundamental role in this analysis and hence they are now studied very intensively since they are also related to famous P. Ho\v{r}ava's proposal
\cite{Horava:2009uw}, for recent review, see \cite{Wang:2017brl}
\footnote{See, for example
\cite{Bergshoeff:2017btm,Bergshoeff:2016lwr,Hartong:2015xda,
Bergshoeff:2015uaa,Andringa:2010it,Bergshoeff:2015ija,
Bergshoeff:2014uea}.}. 

On the other hand asymptotic symmetries of flat space-time were also studied very intensively especially in the context of holography in the flat space-time \cite{Bagchi:2010eg,Bagchi:2012cy,deBoer:2003vf,Arcioni:2003xx,Barnich:2010eb} 
\footnote{For extensive list of references, see for example recent 
work  \cite{Bagchi:2016bcd}.}and 
it was shown that BMS symmetry 
\cite{Bondi:1962px,Sachs:1962zza} has significant meaning in this analysis.
Moreover, remarkable connection between BMS group and Carroll group was recently discovered in \cite{Duval:2014lpa,Duval:2014uva}. Carroll symmetry 
corresponds to the limit of Poincare algebra when the velocity of light
goes to zero \cite{Bacry:1968zf}. There is also very interesting relation between the non-relativistic symmetry  and Carroll symmetry 
\cite{Duval:2014uoa}.

As in case of non-relativistic symmetry
\footnote{\cite{Gomis:2000bd, Danielsson:2000gi,Gomis:2004pw,Kluson:2006xi,Gomis:2016zur,Batlle:2016iel,
Bagchi:2013bga,Bagchi:2015nca,Bagchi:2016yyf}}, Carroll symmetry can be imposed at the level of action for relativistic particle 	\cite{Bergshoeff:2014jla,Bergshoeff:2015wma} where it was shown that the particle possesses trivial dynamics. This analysis was recently extended to another extended objects in string theory, namely fundamental string and $p-$branes in 
\cite{Cardona:2016ytk}. Explicitly, Carroll string and Carroll p-brane were constructed here by taking the Carrollian limits 
of their canonical actions. It was also argued here that the resulting object exhibit trivial dynamics irrespective whether we take different Carroll limits on their actions. 

Despite the fact that the dynamics of Carroll particle, strings and p-branes is trivial it is instructive to apply Carrollian limits of  canonical actions of other objects in string theory, as for example non-BPS Dp-branes
\cite{Sen:1999md,Bergshoeff:2000dq,Kluson:2000iy} which is the goal of present paper. As the first step we consider different Carroll limits in case of unstable p-brane without gauge fields. We analyze the equations of motion that follow from these Carroll unstable p-branes and their tachyon vacuum solutions. It is expected that at the tachyon vacuum with zero electric flux an unstable Dp-brane disappears and we are left with the gas of massless non-interacting particles
\cite{Sen:2000kd,Sen:2003bc,Kluson:2005dr}
 that move in the whole target space-time. We show that this occurs in case of Carroll unstable p-brane as well with one important exception that we have to restrict to the configuration with zero canonical  momenta of the Carroll  particle along  spatial directions where the Carroll contraction has been performed. 
Of course, the dynamics in these directions is again trivial since the 
time derivative of embedding coordinates does not depend on conjugate momenta.

In order to find Carroll non-BPS Dp-brane whose tachyon vacuum state is equivalent to the gas of fundamental string we have to consider Carroll limit of non-BPS Dp-brane action  with world-volume gauge field included.  It turns out that the 
resulting form of the Carroll non-BPS Dp-brane depends on the fact whether we scale gauge field in the same way as embedding coordinates or not. Explicitly, we firstly consider the case where the gauge field are not scaled. Then we found that even in the stringy Carroll limit the Carroll non-BPS Dp-brane at the tachyon vacuum does not have solutions that could be interpreted as fundamental string solutions. For that reason we consider the second possibility where the gauge field scales as well. Now we find that for the stringy Carroll scaling limit non-BPS Dp-brane at the tachyon vacuum possesses fundamental string solutions that can be interpreted as the Carroll string with agreement with the general ideas of the tachyon condensation. On the other hand when we consider 
p-brane Carroll scaling limit on non-BPS Dp-brane we find that this unstable Dp-brane has solution of its tachyon vacuum equations of motion that can be interpreted as Carroll fundamental string that has vanishing momenta along directions where the Carroll limit was taken and also embedding coordinates
along these directions do not depend on spatial coordinate. Of course, all these
solutions are in some sense equivalent since the time evolution of these embedding fields do not depend on conjugate momenta. On the other hand it is an interesting
issue what it is the physical meaning of the solutions of the tachyon equation of motion with non-zero momenta along spatial directions where the Carroll contraction has been performed. 

Let us outline our results. We analyzed different Carroll limits on the world-volume of unstable Dp-brane either with zero or non-zero gauge fields. We discuss the tachyon vacuum solutions and we argued that they correspond to 
the Carroll particle (in case of zero electric flux) or Carroll string (in case of non-zero electric flux) on condition that the momenta along spatial directions where the Carroll limit was performed are equal to zero. It is interesting to compare this result with the fate of the unstable Dp-brane at the tachyon vacuum moving in general space-time, that correspond to the gas of fundamental strings propagating in the whole target space-time. 

It is possible to extend this work in severe directions. For example, since there is a strong similarity between Carroll limit and non-relativistic limit 
we would like to perform non-relativistic limit in case of an unstable Dp-brane, following recent interesting paper \cite{Batlle:2016iel}. This work is currently in progress. 

This paper is organized as follows. In the next section (\ref{second}) we analyze Carroll limit on the world-volume of unstable Dp-brane. Then in section 
(\ref{third}) we extend this analysis to the case of unstable Dp-brane with non-zero gauge field.

\section{Carroll Limit of Non-BPS Dp-brane}\label{second}
\subsection{Stringy Carroll Limit}
In this section we consider stringy Carroll limit of non-BPS Dp-brane
action without gauge fields. Let us start with an  action for unstable Dp-brane
\footnote{We work in units $2\pi\alpha'=1$.} 
\begin{equation}\label{Actpbrane}
S=-\ttau_p\int d^{p+1}\xi
V(\tT)\sqrt{-\det \bA}  \ , 
\bA_{\alpha\beta}=g_{MN}\partial_\alpha \tx^M\partial_\beta \tx^N+
\partial_\alpha \tT\partial_\beta \tT \ , 
\end{equation}
where $\ttau_p$ is a non-BPS Dp-brane tension, 
$\tx^M,M=0,\dots,D-1$ label position of non-BPS Dp-brane in the target space time with the metric $g_{MN}$. Further, $\xi^\alpha,\alpha=0,\dots,p$ are world-volume
coordinates on non-BPS Dp-brane. Finally $T$ is the tachyon field and $V(T)$ is its potential when we presume that $V$ is even function with 
three extremes, where $T_0=0$ is unstable maximum with $V(T_0)=0$ while 
$\pm T_{min}$ are global minima of the potential with $V(\pm T_{min})=0$ 
\cite{Sen:1999md,Bergshoeff:2000dq,Kluson:2000iy}, for review and extensive list of references, see  \cite{Sen:2004nf}.

 In order to formulate Carroll limit 
we have to proceed to the Hamiltonian formalism of the action (\ref{Actpbrane}). From (\ref{Actpbrane}) we find 
conjugate momenta
\begin{eqnarray}
\tp_M&=&-\ttau_p g_{MN}\partial_\alpha \tx^N (\bA^{-1})^{\alpha 0}
\sqrt{-\det \bA} \ , \nonumber \\
\tp_T&=&-\ttau_p\partial_\alpha T (\bA^{-1})^{\alpha
0}\sqrt{-\det \bA} \ . \nonumber \\
\end{eqnarray}
Using these relations it is easy to see that the bare Hamiltonian 
is equal to zero
\begin{eqnarray}
H_B=\int d^p \xi (\tp_M\partial_0 \tx^M+\tp_T\partial_0 \tT-\mL)=0
\end{eqnarray}
while we have following set of primary constraints
\begin{eqnarray}\label{Hamconstpbrane}
\mH_\tau=\tp_Mg^{MN}\tp_N+\ttau_p^2V^2\det \bA_{ij}\approx 0 \ , \quad  i,j=1,\dots,p \ ,  
\quad 
\mH_i=\tp_M\partial_i \tx^M\approx 0 \ .  \nonumber \\
\end{eqnarray} 
Using these constraints we can write 
 non-BPS Dp-brane canonical action   in the form
\begin{equation}\label{actnonBPS}
S=\int d^{p+1}\xi (\tp_M\partial_0 \tx^M+\tp_T\partial_0 \tT-
\tN\mH_\tau-\tN^i\mH_i) \ , 
\end{equation}
where $\mH_\tau,\mH_i$ are given in (\ref{Hamconstpbrane})
and where $\tN,\tN^i$ are corresponding Lagrange multipliers. 
  
Using (\ref{actnonBPS}) we  find  Carroll limit of non-BPS Dp-brane
action. 
Following  \cite{Cardona:2016ytk} we introduce 
"stringy" Carroll limit defined as:
\begin{eqnarray}\label{pbranestringylimit1}
& &\tx^\mu=\frac{X^\mu}{\omega} \ , \quad 
 \tp_\mu=\omega P_\mu \ , \quad \mu,\nu=0,1 \ , \nonumber \\ 
& &\tx^I=X^I \ , \quad \tp_I=P_I \ , \quad  I=2,\dots,D-1 \ ,  \nonumber \\
& & \tT=T \ , \quad  \tp_T=p_T \ , \nonumber \\ 
& &\tN=\frac{1}{\omega^2}N \ , \quad \tN^i=N^i \ , \quad 
\ttau_p=\omega \tau_p \   \nonumber \\
\end{eqnarray}
and take the limit $\omega\rightarrow \infty$. We also consider
string in flat space-time so that $g_{MN}=\eta_{MN}$. 
As  a result of this limiting procedure we obtain  Carroll non-BPS Dp-brane action 
\begin{eqnarray}\label{nonBPSstringaction}
S&=&\int d^{p+1}\xi (P_\mu \partial_0 X^\mu+
P_I\partial_0 X^I+P_T\partial_0 T-\nonumber \\
&-&N
(P_\mu\eta^{\mu\nu}P_\nu+
\tau_p^2V^2\det \ba_{ij})-N^i(P_\mu\partial_i X^\mu+
P_I\partial_i X^I+P_T\partial_i T))  \ ,  
\nonumber \\
\end{eqnarray}
where
\begin{equation}
\ba_{ij}=\partial_i X^I\partial_j X_I+\partial_i T
\partial_j T  \ , \quad  i,j=1,\dots,p  \ . 
\end{equation}
The equations of motion that we derive from (\ref{nonBPSstringaction})
 have the form 
\begin{eqnarray}\label{nonBPSstringactioneq}
& &\partial_0 X^\mu-2N\eta^{\mu\nu}P_\nu-N^i\partial_i X^I=0 \ , \quad 
-\partial_0 P_\mu+\partial_i (N^i P_\mu)=0 \ , \nonumber \\
& &\partial_0 X^I-N^i\partial_i X^I=0 \ , \quad  -\partial_0 P_I+2\tau_p^2
\partial_i (N V^2 \partial_j X^I (\ba^{-1})^{ij})+
\partial_i (N^i P_I)=0 \ ,  \nonumber \\
& &\partial_0 T-N^i\partial_i T=0 \ ,  \quad -\partial_0 P_T-2N\tau_p^2V\frac{d V}{dT}
\det \ba_{ij}+\partial_i (N^i p_T)=0 \ ,\nonumber \\ 
& &P_\mu\eta^{\mu\nu}P_\nu+
\tau_p^2V^2\det \ba_{ij}=0 \ ,  \quad 
 P_\mu\partial_i X^\mu+
P_I\partial_i X^I+P_T\partial_i T=0 \ . \nonumber \\
\end{eqnarray}
We are interested in the interpretation of the 
 equations of motion when Carroll non-BPS Dp-brane 
 is in its  
tachyon vacuum state defined as 
 $T=T_{min} \ , \frac{dV}{dT}(T_{min})=0 \ , 
V(T_{min})=0$ together with $P_T=0$. In this case the equations 
of motion (\ref{nonBPSstringactioneq}) take the form 
\begin{eqnarray}\label{eq1}
& &\partial_0 X^0+2NP_0-N^i\partial_i X^0=0 \ , \quad 
\partial_0 X^1-2NP_1-N^i\partial_i X^1=0 \ , \nonumber \\
& &-\partial_0 P_0+\partial_i (N^iP_0)=0 \ , \quad 
-\partial_0 P_1+\partial_i (N^i
P_1)=0 \ , \nonumber \\
& &\partial_0 X^I-N^i\partial_i X^I=0 \ , \quad -\partial_0 P_I+
\partial_i (N^i P_I)=0 \ , I=2,\dots,D-1 \ ,  \nonumber \\
& &-P_0^2+P_1^2=0 \ ,  \quad 
P_\mu\partial_i X^\mu+
P_I\partial_i X^I=0 \ . \nonumber \\
\end{eqnarray}
It is well known that the non-BPS Dp-brane at its tachyon vacuum 
with zero electric flux is equivalent to the gas of massless particles 
that propagate in the whole target space-time \cite{Kluson:2016tgp}.
Now we would like to see whether this interpretation holds in case of the 
Carroll non-BPS Dp-brane at the  tachyon vacuum.
In other words we would like to see whether 
 these equations of motion have natural 
interpretation as the equations of motion of 
Carroll massless particle \cite{Bergshoeff:2014jla}. To see this in more details
we determine corresponding action starting with an action for massless
relativistic particle 
\begin{equation}\label{Smassparticle}
S=\int d\tau [\tK_M\partial_\tau \tZ^M-\lambda_\tau \tK_M \eta^{MN}
\tK_N] \ . 
\end{equation}
Following \cite{Bergshoeff:2014jla} we introduce particle Carroll limit
as
\begin{eqnarray}
& &Z^0=\frac{t}{\omega} \ , \quad \tK_0=-\omega E \ ,  \quad  \lambda_\tau=\frac{e}{\omega^2} \ , \nonumber \\
& & \tK_I=K_I \ , \quad  
\tZ^I=Z^I \ ,  \quad I=1,\dots, D-1 \ . \nonumber \\
\end{eqnarray}
In this limit the action (\ref{Smassparticle})
 has the form 
\begin{equation}
S=\int d\tau (-\partial_\tau tE+K_I\partial_\tau Z^I+eE^2)
\end{equation}
with corresponding  equations of motion 
\begin{eqnarray}\label{parteofmot}
& &-\partial_\tau t+2eE=0 \ , \quad  \partial_\tau E=0 \ , \nonumber \\
& &\partial_\tau Z^I=0 \ , \quad \partial_\tau K_I=0 \ , \nonumber \\
& & E^2=0 \ . 
\nonumber \\
\end{eqnarray}
First of all, comparing (\ref{eq1}) with the last equation 
in (\ref{parteofmot}) we immediately see that in order 
the solutions of the equations of motion (\ref{eq1}) 
have massless particle interpretation we have to demand that
$P_1=0$. Then following \cite{Sen:2000kd,Kluson:2016tgp} we now 
propose an ansatz for the solution of the equation
of motion (\ref{eq1})  
\begin{eqnarray}\label{partansat}
& & P_0(\xi^0)|_{\xi^0=\tau}=-E(\tau)  \ , \quad X^0(\xi^0)|_{\xi^0=\tau}=t(\tau) \ , 
 \nonumber \\
& &  X^1(\xi^0)|_{\xi^0=\tau}=Z^1(\tau) \ , \nonumber \\
& & X^I(\xi^0)|_{\xi^0=\tau}=Z^I(\tau) \ , \quad
 P_I(\xi^0)|_{\xi^0=\tau}=K_I(\tau) \ ,  \quad 
 I=2,\dots,D-1 \ ,  \nonumber \\
& & N(\xi^0)|_{\xi^0=\tau}=e(\tau) \ , \quad 
 N^i=0 \ , \quad  i=1,\dots,p  \ . \nonumber \\
\end{eqnarray}
We see that this is solution of the equations of motion (\ref{eq1}) on condition that $Z^I,Z^0,K_0,Z^i$ obey the equation of motion 
(\ref{parteofmot}) with condition $P_i=0$.

\subsection{p-brane Carroll Limit} 
In previous section we introduced  Carroll limit $\grave{a} \ la \ string$ when 
we scale first two target space coordinates and conjugate momenta. 
We can also define Carroll limit $\grave{a} \ la \ particle$ when we only scale
time coordinate \cite{Cardona:2016ytk}. 
In this section we consider  Carroll limit $\grave{a} \ la \ p-brane$  that is defined as \cite{Cardona:2016ytk}
\begin{eqnarray}\label{pbranescal1}
& &\tx^\mu=\frac{X^\mu}{\omega} \ , \quad  p_\mu=\omega P_\mu \ , \quad  \mu=0,\dots, p   \ , \nonumber \\
& & \tx^I=X^I \ , \quad p_I=P_I \ , \quad  I=p+1,\dots,D-1 \ ,  \nonumber \\ 
& & \tN=\frac{1}{\omega^2}N \ , \quad 
 \tN^i=N^i  \ , \quad 
\ttau_p=\omega \tau_p  \ , \nonumber \\
& &\tT=T \ , \quad  p_T=P_T \ . \nonumber \\
\end{eqnarray}
The difference between stringy and p-brane Carroll limit is in 
the number of scalar modes that are scaled. Clearly we could define
$q-brane$ Carroll limit for any $0\leq q \leq p$ where $q=0$ corresponds to Carroll particle limit, $q=1$ corresponds to stringy Carroll limit and so on.
Performing the scaling limit (\ref{pbranescal1}) in the action (\ref{actnonBPS})
we obtain p-brane Carroll non-BPS Dp-brane action in the form 
\begin{eqnarray}\label{nonBPSstringactionplimit}
S&=&\int d^{p+1}\xi (P_\mu \partial_0 X^\mu+
P_I\partial_0 X^I+P_T\partial_0 T-\nonumber \\
&-&N
(P_\mu\eta^{\mu\nu}P_\nu+
\tau_p^2V^2\det \ba_{ij})-N^i(P_\mu\partial_i X^\mu+
P_I\partial_i X^I+P_T\partial_i T))  \ ,  
\nonumber \\
\end{eqnarray}
where
\begin{equation}
\ba_{ij}=\partial_i X^I\partial_j X_I+\partial_i T
\partial_j T  \ , \quad i,j=1,\dots,p  \ , \quad 
 I,J=p+1,\dots,D-1 \ . 
\end{equation}
The equations of motion that we derive from (\ref{nonBPSstringactionplimit}) at the tachyon 
vacuum 
have the form 
\begin{eqnarray}\label{nonBPSplimiteom}
& &\partial_0 X^\mu-2N\eta^{\mu\nu}P_\nu-N^i\partial_i X^I=0 \ , \quad 
-\partial_0 P_\mu+\partial_i (N^i P_\mu)=0 \ , \nonumber \\
& &\partial_0 X^I-N^i\partial_i X^I=0 \ , \quad -\partial_0 P_I+
\partial_i (N^i P_I)=0 \ ,  \nonumber \\
& &P_\mu\eta^{\mu\nu}P_\nu
=0 \ , \quad 
P_\mu\partial_i X^\mu+
P_I\partial_i X^I+P_T\partial_i T=0 \ . \nonumber \\
\end{eqnarray}
We can again presume that at the tachyon vacuum non-BPS Dp-brane
disappears and we should find collection of Carroll particles.
From the first equation on the last line above we find that this is possible
when $P_m=0 \ , m=1,\dots,p$. 

 In order to support this arguments let us consider following ansatz
\begin{eqnarray}\label{partansat2}
& &P_0(\xi^0)|_{\xi^0=\tau}=-E(\tau)  \ , \quad X^0(\xi^0)|_{\xi^0=\tau}=t(\tau) \ , 
\nonumber \\
& &
X^m(\xi^0)|_{\xi^0=\tau}=Z^m(\tau) \ , \quad m=1,\dots,p \ , \nonumber \\
& &X^I(\xi^0)|_{\xi^0=\tau}=Z^I(\tau) \ , \quad 
P_I(\xi^0)|_{\xi^0=\tau}=K_I(\tau) \ , \quad 
I=2,\dots,D-1 \ ,  \nonumber \\
& & N(\xi^0)|_{\xi^0=\tau}=e(\tau) \ , \quad N^i=0 \ .
 \nonumber \\
\end{eqnarray}
It is easy to see that (\ref{partansat2}) solve the equations
of motion  (\ref{nonBPSplimiteom})  on 
condition that $Z^I,Z^0,Z^m,K_m=0$ obey the equations of motion 
(\ref{parteofmot}). In other words, the solutions of the equation of motion of 
Carroll non-BPS Dp-brane at the tachyon vacuum describes massless Carroll
particles with vanishing momenta $K_m=0$. Note that they are the momenta along spatial directions where the Carroll limit has been performed.  
\section{Carroll Limit of Non-BPS Dp-Brane with Gauge Field}\label{third}
We would like to see whether it is possible to have Carroll limit for non-BPS Dp-brane that at the tachyon vacuum has solution of the equations of motion corresponding to the fundamental string. As we know from the analysis of unstable Dp-brane the electric field on the world-volume of non-BPS Dp-brane is crucial
\cite{Sen:2000kd,Kluson:2016tgp}.
 For that reason we have to consider Carroll limit for non-BPS Dp-brane with the gauge field. Recall that such an action has the form 
\begin{eqnarray}\label{actnongauge}
S&=&-\ttau_p\int d^{p+1}\xi
V(\tT)\sqrt{-\det \bA_{\alpha\beta}} 
 \ ,  \nonumber \\
 & &\bA_{\alpha\beta}=g_{MN}\partial_\alpha \tx^M\partial_\beta \tx^N+
 \tF_{\alpha\beta}+\partial_\alpha \tT\partial_\beta \tT \ , \quad  \tF_{\alpha\beta}=
 \partial_\alpha \tA_\beta-\partial_\beta \tA_\alpha \ .  \nonumber \\
 \end{eqnarray}
In order to define Carroll limit we have to find canonical  form of the action which means that we have to find corresponding Hamiltonian. 
Explicitly, from (\ref{actnongauge}) we find 
\begin{eqnarray}\label{momnongauge}
\tp_M&=&-\ttau_pV g_{MN}\partial_\beta \tx^N (\bA^{-1})^{\beta 0}_S
\sqrt{-\det \bA} \ , \nonumber \\
\tp_T&=&-\ttau_pV \partial_\beta \tT (\bA^{-1})^{\beta 0}_S\sqrt{-\det \bA} \ , 
\nonumber \\
\tpi^i&=&
-\ttau_pV (\bA^{-1})^{i0}_A \sqrt{-\det \bA} \ , \quad 
\tpi^0\approx 0 \ , \nonumber \\
\end{eqnarray}
where
\begin{equation}
(\bA^{-1})^{\alpha\beta}_S=\frac{1}{2}((\bA^{-1})^{\alpha\beta}+
(\bA^{-1})^{\beta\alpha}) \ , \quad 
(\bA^{-1})^{\alpha\beta}_A=\frac{1}{2}((\bA^{-1})^{\alpha\beta}-
(\bA^{-1})^{\beta\alpha}) \ .
\end{equation}
Using (\ref{momnongauge}) we easily find that the bare
Hamiltonian is equal to
\begin{equation}
H_B=\int d^p\xi \pi^i\partial_i A_0
\end{equation}
while we have $p+1$ primary constraints
\begin{eqnarray}
\mH_\tau&=&\tp_M g^{MN}\tp_N+\tp_T^2+\tpi^i \bA^S_{ij}\tpi^j+\ttau_p^2V^2
\det \bA_{ij}\approx 0 \ , 
\nonumber \\
\mH_i&=&\tp_M \partial_i \tx^M+\tp_T\partial_i \tT+\tF_{ij}\tpi^j 
\nonumber \\
\end{eqnarray}
so that the canonical action is equal to
\begin{eqnarray}\label{firstordernongauge}
S=\int d^{p+1}\xi (\tp_M\partial_0 \tx^M+\tp_T\partial_0
\tT+\tpi^i\partial_0 \tA_i-\tN \tmH_\tau-\tN^i\mH_i-\tpi^i\partial_i \tA_0) \ . 
\nonumber \\
\end{eqnarray}
As in the first section we firstly  consider "stringy" Carroll limit 
\begin{eqnarray}\label{pbranestringylimitgauge1}
& &\tx^\mu=\frac{X^\mu}{\omega} \ , \quad 
\tp_\mu=\omega P_\mu \ , \nonumber \\ 
& &\tx^I=X^I \ ,  \quad \tp_I=P_I \ , \quad  I=2,\dots,D-1 \ ,  \nonumber \\
& & \tT=T \ ,  \quad \tp_T=p_T \ , \nonumber \\ 
& &\tN=\frac{1}{\omega^2}N \ , \quad \tN^i=N^i \ , \quad 
\ttau_p=\omega \tau_p \  , \nonumber \\
& &\tA_i=A_i \ , \quad \tpi^i=\pi^i \ .
\end{eqnarray}
Then in the limit $\omega\rightarrow \infty$
we obtain action for stringy"Carroll non-BPS Dp-brane 
in the form
$\omega\rightarrow \infty$
\begin{eqnarray}\label{SCarrollgauge1}
S&=&\int d^{p+1}\xi (P_\mu \partial_0 X^\mu+P_I\partial_0 X^I+p_T\partial_0 
T+\pi^i\partial_0 A_i-\nonumber \\
&-&N(P_\mu \eta^{\mu\nu}P_\nu +\tau_p^2 V^2
\det \ba_{ij})-N^i(P_\mu \partial_i X^\mu+
P_I\partial_i X^I+p_T\partial_i T+F_{ij}\pi^j) -\pi^i\partial_i A_0) \ , 
\nonumber \\
\end{eqnarray}
where
\begin{equation}
\ba_{ij}=\partial_i X^I\partial_j X_I+\partial_i T\partial_jT+
F_{ij}  \ . 
\end{equation}
From (\ref{SCarrollgauge1}) we determine following equations of motion
\begin{eqnarray}\label{SCarrollgauge1eqm}
& &\partial_0 X^\mu-2N\eta^{\mu\nu}P_\nu-N^i\partial_i X^\mu=0 \ ,
\quad 
-\partial_0 P_\mu+\partial_i (N^i P_\mu)=0 \ , \nonumber \\
& &\partial_0 X^I-N^i\partial_i X^I=0 \ , \quad 
-\partial_0 P_I+2\tau_p^2 \partial_i[N V^2 \delta_{IJ}
\partial_j X^J (\ba^{-1})^{ji}_S\det \ba]+\partial_i (N^i P_I)=0 \ , 
\nonumber \\
& &\partial_0 T-N^i\partial_i T=0 \ ,\quad 
-\partial_0 p_T -2N\tau_p^2 V\frac{dV}{dT}\det \ba_{ij}+
\partial_i (N^i p_T)=0 \ , \nonumber \\
& &\partial_0 A_i-N^j F_{ji}-\partial_i A_0=0 \ , \quad 
-\partial_0 \pi^i+\partial_j (N^j\pi^i)-\partial_j (N^i\pi^j)=0 \ ,
\quad  \partial_i \pi^i=0  \ , 
\nonumber \\
& & 
 P_\mu \eta^{\mu\nu}P_\nu +\tau_p^2 V^2
\det \ba_{ij}=0 \ , \quad 
P_\mu \partial_i X^\mu+
P_I\partial_i X^I+p_T\partial_i T+F_{ij}\pi^j=0 \ . \nonumber \\
\end{eqnarray}
These equations of motion simplify considerably at the tachyon 
vacuum  
\begin{eqnarray}\label{SCarrollgauge1eqmvac}
& &\partial_0 X^\mu-2N\eta^{\mu\nu}P_\nu-N^i\partial_i X^\mu=0 \ ,
\quad -\partial_0 P_\mu+\partial_i (N^i P_\mu)=0 \ , \nonumber \\
& &\partial_0 X^I-N^i\partial_i X^I=0 \ , \quad 
-\partial_0 P_I+\partial_i (N^i P_I) \ , 
\nonumber \\
& &\partial_0 A_i-N^j F_{ji}-\partial_i A_0=0 \ , \quad 
-\partial_\tau \pi^i+\partial_j (N^j\pi^i)-\partial_j (N^i\pi^j)=0 \ ,
\nonumber \\
& &\partial_i \pi^i=0 \ , \nonumber \\ 
& & P_\mu \eta^{\mu\nu}P_\nu=0 \ , \quad 
P_\mu \partial_i X^\mu+
P_I\partial_i X^I+p_T\partial_i T+F_{ij}\pi^j=0 \ . \nonumber \\
\end{eqnarray}
We would like to see whether these equations of motion 
have solutions that can be interpreted as the Carroll string moving 
in the target space-time. To see this we have to determine Carroll
limit of fundamental string action, following 
\cite{Cardona:2016ytk}. We start with the action 
\begin{equation}
S=-\ttau_F \int d\tau d\sigma \sqrt{-\det G_{\alpha\beta}} \ , \quad
G_{\alpha\beta}=G_{MN}\partial_\alpha \tZ^M\partial_\beta \tZ^N \ . 
\end{equation} 
Then we obtain
\begin{equation}
\tK_M=-\tau_F g_{MN}\partial_\beta \tZ^N G^{\beta\tau} \sqrt{-\det G_{\alpha
\beta}}
\end{equation}
so that it is easy to see that we have two primary constraints
\begin{equation}
\mK_\tau=\tK_M g^{MN}\tK_N+\ttau_F^2 \partial_\sigma \tZ^M \partial_\sigma \tZ^N
g_{MN}\approx 0 \ ,  \quad 
\mK_\sigma=\tK_M \partial_\sigma \tZ^M \approx 0 
\end{equation}
and hence  the canonical  action  has the form 
\begin{equation}
S=\int d\tau d\sigma (\tK_M\partial_\tau \tZ^M-\tlambda_\tau \mK_\tau -
\tlambda_\sigma \mK_\sigma) \ ,
\end{equation}
where $\tlambda_\tau,\tlambda_\sigma$ are Lagrange multipliers corresponding to the primary constraints $\mK_\tau,\mK_\sigma$. Now we take stringy
Carroll limit when 
\begin{eqnarray}
& &\tZ^\mu=\frac{Z^\mu}{\omega} \ , \quad 
\tK_\mu=\omega K_\mu \ , \quad  \mu,\nu=0,1 \ , 
\nonumber \\
& &\tZ^I=Z^I \ , \quad \tK_I=K_I \ , \quad  I=2,\dots,D-1 \ , \nonumber \\
& & \ttau_F=\omega \tau \ , \quad  \tlambda_\tau
=\frac{\lambda_\tau}{\omega^2} \ , \quad  \tlambda_\sigma=\lambda_\sigma
\end{eqnarray}
so that the action has the form 
\begin{equation}
S=\int d^2\sigma (K_\mu \partial_\tau Z^\mu+K_I\partial_\tau Z^I-
\lambda_\tau (K_\mu \eta^{\mu\nu}K_\nu+\tau_F^2 \partial_\sigma Z^I
\partial_\sigma Z_I)-\lambda_\sigma (K_\mu \partial_\sigma Z^\mu
+K_I\partial_\sigma Z^I)) \ . 
\end{equation}
The equations of motion that follow from this action have the form 
\begin{eqnarray}\label{eqCarrollstring}
& &\partial_\tau Z^\mu-2\lambda_\tau \eta^{\mu\nu}K_\nu-\lambda_\sigma 
\partial_\sigma Z^\mu=0 \ , \quad 
-\partial_\tau K_\mu+\partial_\sigma(\lambda_\sigma K_\mu)=0 \ , \nonumber \\
& &\partial_\tau Z^I-\lambda_\sigma \partial_\sigma Z^I=0 \ , \quad 
-\partial_\tau Z_I+2\tau_F^2\partial_\sigma(\lambda_\tau 
\partial_\sigma Z_I)+\partial_\sigma (\lambda_\sigma K_I)= 0 \  ,
\nonumber \\
& &K_\mu \eta^{\mu\nu}K_\nu+\tau_F^2 \partial_\sigma Z^I
\partial_\sigma Z_I=0 \ , \quad 
K_\mu \partial_\sigma Z^\mu
+K_I\partial_\sigma Z^I=0  \ . \nonumber \\
\end{eqnarray}
Comparing (\ref{eqCarrollstring}) with (\ref{SCarrollgauge1eqmvac}) 
we immediately obtain that there is no way how solutions of the equations
(\ref{eqCarrollstring}) could be related to the solution of the equation 
(\ref{SCarrollgauge1eqmvac}) due to the fact that the later contains term proportional to the string tension while such a term is absent in (\ref{SCarrollgauge1eqmvac}). In order to resolve this issue we propose 
Carroll  scaling limit when the gauge field scales as well.
\subsection{Modified Stringy Carroll  Scaling limit}
We define this new Carroll scaling limit as  
\begin{eqnarray}\label{secscall}
& &\tx^\mu=\frac{X^\mu}{\omega} \ , \quad  \tp_\mu=\omega P_\mu \ ,  \quad 
\tx^I=X^I \ , \quad \tp_I=P_I \ , \nonumber \\ 
& & \tA_i=\frac{A_i}{\omega} \ , \quad  \tpi^i=\omega \pi^i \ , \quad  \tA_0=\frac{1}{\omega}
A_0 \ ,  \quad  \tN=\frac{1}{\omega^2}
N \ , \quad \tN^i=N^i  \ , \quad 
\ttau_p=\omega \tau_p  \ , \nonumber \\
& &\tT=T \ , \quad  p_T=P_T \ . \nonumber \\
\end{eqnarray}
Then in the limit $\omega\rightarrow \infty (g_{MN}=\eta_{MN})$ the 
 the action (\ref{firstordernongauge})  has the form 
\begin{eqnarray}\label{stringgaugeaction}
& &S=\int d^{p+1}\xi (P_\mu \partial_0 X^\mu+P_I\partial_0 X^I+p_T\partial_0
T+\pi^i\partial_0 A_i -\pi^i\partial_i A_0-\nonumber \\
& &-N(P_\mu \eta^{\mu\nu}P_\nu +\pi^i\ba_{ij}\pi^j+\tau_p^2V^2
\det \ba_{ij})-N^i(P_\mu \partial_i X^\mu+
P_I\partial_i X^I+p_T\partial_i T+F_{ij}\pi^j)) \ , 
\nonumber \\
\end{eqnarray}
where
\begin{equation}
\ba_{ij}=\partial_i X^I\partial_j X_I+\partial_i T\partial_jT \ . 
\end{equation}
We see the crucial difference with respect to the previous form of the
action
(\ref{SCarrollgauge1}) 
 which is in the presence of the term $\pi^i\ba_{ij}\pi^j$ in the Hamiltonian
constraint that is important for string like interpretation of the 
solutions of the equations of motion at the tachyon vacuum. 
Explicitly, from (\ref{stringgaugeaction}) we obtain following equations
of motion 
\begin{eqnarray}\label{eqgaugebrane}
& &\partial_0 X^\mu-2N\eta^{\mu\nu}P_\nu-N^i\partial_i X^\mu=0 \ ,
\quad
-\partial_0 P_\mu+\partial_i (N^i P_\mu)=0 \ , \quad 
\ , \nonumber \\
& &-\partial_0 P_I
+2\partial_i[N\pi^i \delta_{IJ}\partial_j X^J\pi^j]
+2\tau_p^2 \partial_i[N V^2 \delta_{IJ}
\partial_j X^J (\ba^{-1})^{ji}\det \ba]+\partial_i (N^i P_I)=0 \ , 
\nonumber \\
& &\partial_0 X^I-N^i\partial_i X^I=0  \ , \quad \partial_0 T-N^i\partial_i T=0 \ ,  \nonumber \\
& &  
-\partial_\tau P_T+2\partial_i (N\pi^i\partial_j T\pi^j)+
2\tau^2_p\partial_i[NV^2 \partial_j T (\ba^{-1})^{ji}]
 -2N\tau^2_p V\frac{dV}{dT}\det \ba_{ij}+
\partial_i (N^i p_T)=0 \ , \nonumber \\
& &\partial_0 A_i-N^j F_{ji}-\partial_i A_0-2N \ba_{ij}\pi^j=0 \ , \quad 
-\partial_0 \pi^i+\partial_j (N^j\pi^i)-\partial_j (N^i\pi^j)=0 \ , \quad 
\partial_i \pi^i=0 \ , \nonumber \\ 
& & P_\mu \eta^{\mu\nu}P_\nu +\pi^i\ba_{ij}\pi^j+\tau_p^2V^2
\det \ba_{ij}=0 \ , \quad P_\mu \partial_i X^\mu+
P_I\partial_i X^I+p_T\partial_i T+F_{ij}\pi^j=0 \ . \nonumber \\
\end{eqnarray}
At the tachyon vacuum these equations of motion simplify considerably
and have the form 
\begin{eqnarray}\label{eqptach}
& &\partial_0 X^\mu-2N\eta^{\mu\nu}P_\nu-N^i\partial_i X^\mu=0 \ ,
\quad 
-\partial_0 P_\mu+\partial_i (N^i P_\mu)=0 \ , \nonumber \\
& &\partial_0 X^I-N^i\partial_i X^I=0 \ , \quad 
-\partial_0 P_I
+2\partial_i[N\pi^i \delta_{IJ}\partial_j X^J\pi^j]
+\partial_i (N^i P_I)=0 \ , 
\nonumber \\
& &-\partial_0 \pi^i+\partial_j (N^j\pi^i)-\partial_j (N^i\pi^j)=0 \ ,
\quad 
\partial_i \pi^i=0 \ , \nonumber \\ 
& & P_\mu \eta^{\mu\nu}P_\nu +\pi^i\ba_{ij}\pi^j=0 \ , \quad P_\mu \partial_i X^\mu+
P_I\partial_i X^I+p_T\partial_i T+F_{ij}\pi^j=0 \ . \nonumber \\
\end{eqnarray}
In order to find string like solutions we proceed in the similar way as in 
\cite{Kluson:2016tgp}.  We
introduce a projector 
\begin{equation}
\triangle^i_{ \ j}=\delta^i_j-\frac{\gamma_{jk}\pi^k \pi^i}{
	\pi^m \gamma_{mn}\pi^n} \ , \quad \gamma_{ij}=\partial_i X^I
\partial_j X_I \ 
\end{equation}
that obeys the property
\begin{equation}
\triangle^i_{ \ j}\pi^j=0 \ , \quad 
\triangle^i_{ \ j}\triangle^j_{ \ k}=\triangle^i_{ \ k} \ . 
\end{equation}
In other words $\triangle^i_{ \ j}$ is the projector on the space
transverse to $\pi^i$. Then we split $N^i$ as
\begin{equation}
N^i=N^j\delta_j^i=\triangle^i_{ \ j}N^j+\pi^i\frac{\pi^k \gamma_{kj}}
{\pi^m \gamma_{mn}\pi^n}N^j=N_{\bot}^i+\pi^i N_{II} \ , 
\end{equation}
where by definition $N_{\bot}^i \gamma_{ij}\pi^j=0$. We again introduce
dimensionless $\tpi^i$ defined as $\pi^i=\tau_p \tpi^i$ and search for the solution when $\pi^i$ is constant. Then the last equation in (\ref{eqptach}) is automatically 
obeyed while the previous one implies
\begin{eqnarray}
\tau_p \partial_\sigma N_{\bot}\pi^i-
\tau_p \partial_\sigma N^i_{\bot}=0 \ . \nonumber \\
\end{eqnarray}
Since this equation has to hold for any $\pi^i$ we have to demand that 
\begin{equation}
\partial_\sigma N^i_{\bot}=0 \ . 
\end{equation}
Then it is easy to see that the equations of motion 
(\ref{eqptach})
 have the form
\begin{eqnarray}\label{eqCarrollstringeq1}
& &\partial_0 X^\mu-2N\eta^{\mu\nu}P_\nu-N_{II}\tau_p \partial_\sigma X^\mu=0 \ ,
\nonumber \\
& & \partial_0 P_\mu+\tau_p\partial_\sigma (N_{II} P_\mu)=0 \ , \nonumber \\
& &\partial_0 X^I-\tau_p N_{II}\partial_\sigma X^I=0 \ , \nonumber \\
& &-\partial_0 P_I+2\tau_p^2 \partial_\sigma
[Ng_{IJ}\partial_\sigma X^J]+
\tau_p \partial_\sigma (N_{II}  P_I)=0 \ 
\end{eqnarray}
that correspond to the equations of motion of the 
 Carroll string (\ref{eqCarrollstring}) 
when we identify 
$\lambda_\sigma=\tau_p N_{II}$ and $\lambda_\tau \tau^2_F=
\tau^2_p N$ (More precise identification will be given in the next section). In other words tachyon vacuum solutions of the stringy Carroll
non-BPS Dp-brane can be interpreted as Carroll fundamental string which is 
in agreement with the general ideas of the tachyon condensation on unstable Dp-brane. 
Remarkable property of this solution is that it does not depend on all world-volume coordinates of non-BPS Dp-brane but it only depends on 
$\sigma$, where $\sigma$ is defined by the orientation of the electric
flux on the world-volume of non-BPS Dp-brane at the tachyon vacuum. We mean
that this is a natural result if we recognize that it is believed that at the tachyon
vacuum the non-BPS Dp-brane disappears. Further, note that the localization of
the electric flux on the world-volume of non-BPS Dp-brane does not have physical
meaning when the full world-volume diffeomorphism invariance is preserved. Of course, this analysis is based on "stringy" Carroll limit when we perform 
Carroll scaling limit on two coordinates that coincide with 
the Carroll scaling limit of canonical string action. More interesting situation occurs when we consider 
 Carroll limit $\grave{a} \ la \ p-brane$ when we perform contraction in all $p-$spatial coordinates. 

Before we proceed to the $p-$Brane Carroll Limit let us solve  equations
of motion  (\ref{eqCarrollstringeq1}) in different gauges. As the first case we consider  static gauge defined as
\begin{equation}
X^0=\tau \ , X^1=\sigma \ . 
\end{equation}
In this gauge  the first equation in (\ref{eqCarrollstringeq1}) implies
\begin{equation}
N=-\frac{1}{2P_0} \ , \quad  \tau_p N_{II}=\frac{P_1}{P_0} \ 
\end{equation}
that is of course valid on condition that $P_0\neq 0$. 
To proceed further note that  $P_1$ can be expressed using the spatial  diffeomorphism constraint as
\begin{equation}
P_1=-P_I\partial_\sigma X^I \
\end{equation}
while $P_0$ can be expressed from the Hamiltonian constraint as
\begin{equation}
P_0^2=(P_I\partial_\sigma X^I)^2+\tau_p^2 \partial_\sigma X^I\partial_\sigma X_I
 \ . 
\end{equation}
Now the condition that $P_0\neq 0$ implies that $\partial_\sigma X^I\neq 0$
 so that the remaining equations of motion have the form 
 \begin{eqnarray}\label{remeqm}
& & \partial_0 X^I+ \frac{P_J\partial_\sigma X^J}{P_0}\partial_\sigma X^I=0 \ , 
 \nonumber \\
& & \partial_0 P_I-\tau_p^2\partial_\sigma\left[\frac{1}{P_0}\partial_\sigma X_I\right]
-\partial_\sigma \left[\frac{P_J\partial_\sigma X^JP_I}{P_0}\right]=0 \ . 
 \nonumber \\
 \end{eqnarray}
We observe that when we introduce Hamiltonian density $\mH_{red}$ on the 
reduced phase space defined as
\begin{equation}
\mH_{red}=-P_0=\sqrt{(P_I\partial_\sigma X^I)^2+\tau_p^2
\partial_\sigma X^I\partial_\sigma X_I} \ , \quad  H_{red}=\int d\sigma \mH_{red}
\end{equation} 
 we can write the equations of motion (\ref{remeqm}) into the form 
 \begin{eqnarray}
 \partial_0 X^I=\pb{X^I,H_{red}}_{red}  \ , \quad 
 \partial_0 P_I=\pb{P_I,H_{red}}_{red} \ , 
 \nonumber \\
 \end{eqnarray}
where  we defined reduced Poisson brackets 
\begin{equation}
\pb{X,Y}_{red}=\int d\sigma \left(\frac{\partial X}{\partial 
X^I(\sigma)}\frac{\partial Y}{\partial P_I(\sigma)}-
\frac{\partial Y}{\partial 
	X^I(\sigma)}\frac{\partial X}{\partial P_I(\sigma)}\right) \ . 
 \end{equation}
It is easy to see that the time independent configuration corresponds to the situation when $P_I=0$ that solves the equation of motion for $P_I$
on condition when 
\begin{equation}\label{parXk}
\frac{\partial_\sigma X^I}{\sqrt{\partial_\sigma X^J\partial_\sigma X_J}}=k^I \ , 
k^I=\mathrm{const} \ , 
\end{equation}
where $k^I$ obey the equation $k^Ik_I=1$ so that (\ref{parXk}) has solution
$X^I=k^I\sigma$. We should stress that imposing the static gauge
we implicitly work with the infinite extended string so that this configuration
corresponds to the straight infinite Carroll string with orientation in the
target space-time determined by parameters $k^I$. 

As the second case  we consider the solution in the  conformal gauge defined 
by the condition 
\begin{equation}
N_{II}=0 \ , \quad N=\frac{1}{2} \ . 
\end{equation}
where the equations (\ref{eqCarrollstringeq1}) simplify considerably 
\begin{eqnarray}\label{eqstringconformal}
& &\partial_0 X^\mu-\eta^{\mu\nu}P_\nu=0 \ ,  \quad 
\partial_0 P_\mu=0 \ , \nonumber \\
& &\partial_0 X^I=0 \ , \quad 
-\partial_0 P_I+\tau_p^2\partial_\sigma^2 X_I=0 \ . 
\nonumber \\
& &\eta^{\mu\nu}P_\mu P_\nu+\partial_\sigma X^I\partial_\sigma X_I=0 \ , 
\quad P_\mu \partial_\sigma X^\mu+P_I\partial_\sigma X^I=0 \ . 
\nonumber \\
\end{eqnarray}
The simplest solution corresponds to $X_I=c_I$ where $c_I$ are constants. 
Then the second equation on the second line in (\ref{eqstringconformal}) implies that $P_I=p_I(\sigma)$ while from the Hamiltonian constraint we find 
$P_0= P_1$ where we have chosen the positive branch without lost of generality. 
Then the spatial diffeomorphism constraint implies
\begin{equation}
X^0+X^1=f(\tau) 
\end{equation} 
while the first equation on the first line in (\ref{eqstringconformal}) gives 
(for $P_1=P_0$)
\begin{equation}
\partial_0 (X^0+X^1)=0  
\end{equation}
and hence $ X^0+X^1=0$, where we choosen the integration constant to be equal to 
zero without lost of generality. Finally 
since $P_0=p_0(\sigma)$ we find that 
\begin{equation}
\partial_0(X^1-X^0)=2p_0(\sigma) 
\end{equation}
that implies $X^1-X^0=2p_0(\sigma)\tau$ which implies in the end
\begin{equation}
X^1=p_0\tau \ , X^0=-2p_0\tau \ . 
\end{equation}
In summary, we can find different solutions of the Carroll string equations of motion whose behavior depend on the gauge we impose and on the profile of the initial data. 
 
\subsection{$p-$Brane Carroll Limit}
We studied $p-brane$ Carroll limit in section (\ref{second}). In this section we generalize this analysis to the case of non-zero gauge fields
and propose following $p-$brane Carroll limit 
\begin{eqnarray}\label{pbranescal}
& & \tx^\mu=\frac{X^\mu}{\omega} \ , \quad  \tp_\mu=\omega P_\mu 
 \ , \quad  \mu=0,\dots, p  \ ,  \nonumber \\
& &\tx^I=X^I \ , \quad \tp_I=P_I \ , \quad  I=p+1,\dots,D-1 \ ,  \nonumber \\ 
& &\tA_i=\frac{A_i}{\omega} \ ,  \quad \tpi^i=\omega \pi^i \ , \quad  \tA_0=\frac{1}{\omega}
A_0 \ , \nonumber \\
& & \tN=\frac{1}{\omega^2}
N \ , \quad  \tN^i=N^i  \ , \quad 
\ttau_p=\omega \tau_p  \ , \nonumber \\
& &\tT=T \ , \quad  p_T=P_T \ . \nonumber \\
\end{eqnarray}
in the action (\ref{firstordernongauge}) . Then in 
the limit
$\omega\rightarrow \infty$ we obtain following p-brane Carroll 
non-BPS Dp-brane action
\begin{eqnarray}\label{nongaugepbrane}
& &S=\int d^{p+1}\xi (P_\mu \partial_0 X^\mu+P_I\partial_0
 X^I+p_T\partial_0 
T+\pi^i\partial_0 A_i -\pi^i\partial_i A_0-\nonumber \\
& &-N(P_\mu \eta^{\mu\nu}P_\nu +\pi^i\ba_{ij}\pi^j+\tau_p^2V^2
\det \ba_{ij})-N^i(P_\mu \partial_i X^\mu+
P_I\partial_i X^I+p_T\partial_i T+F_{ij}\pi^j)) \ , 
\nonumber \\
\end{eqnarray}
where
\begin{equation}
\ba_{ij}=\partial_i X^I\partial_j X_I+\partial_i T\partial_jT \ . 
\end{equation}
It is easy to determine equations of motion from the action 
(\ref{nongaugepbrane}) that have formally the same form as the 
equations of motion (\ref{eqgaugebrane}) where now $\mu=0,\dots,p,
I,J=p+1,\dots,D-1$. We are again interested for the analysis of 
these equations of motion at the tachyon vacuum where they 
have the form 
\begin{eqnarray}\label{nonBPS}
& &\partial_0 X^\mu-2N\eta^{\mu\nu}P_\nu-N^i\partial_i X^\mu=0 \ ,
\quad 
-\partial_0 P_\mu+\partial_i (N^i P_\mu)=0 \ , \nonumber \\
& &\partial_0 X^I-N^i\partial_i X^I=0 \ , \quad 
-\partial_0 P_I
+2\partial_i[N\pi^i \delta_{IJ}\partial_j X^J\pi^j]
+\partial_i (N^i P_I)=0 \ , 
\nonumber \\
& &
\partial_0 A_i-2N \ba_{ij}\pi^j-N^jF_{ji}=0 \ ,\quad 
-\partial_0 \pi^i+\partial_j (N^j\pi^i)-\partial_j (N^i\pi^j)=0 \ ,
\quad 
\partial_i \pi^i=0 \ , \nonumber \\ 
& & P_\mu \eta^{\mu\nu}P_\nu +\pi^i\ba_{ij}\pi^j=0 \ , \quad P_\mu \partial_i X^\mu+
P_I\partial_i X^I+F_{ij}\pi^j=0 \ , \nonumber \\
\end{eqnarray}
where we again stress $\mu,\nu=0,\dots,p, I,J=p+1,\dots,D-1$.
From the Hamiltonian constraint given in (\ref{nonBPS}) we see
that in order to have string like solution we have to demand that $P_m=0$ for $m=2,\dots,p$. Further, since $\ba_{ij}=
\partial_i X^I\partial_j X_I$ for $I=p+1,\dots,D-1$ we see 
that string solution has to be chosen such that $Z^m,m=2,\dots,p$ do not depend on $\sigma$.  Then we can formally proceed as in previous section. Firstly 
we introduce projector 
\begin{equation}
\triangle^i_{ \ j}=\delta^i_j-\frac{\gamma_{jk}\pi^k \pi^i}{
	\pi^i \gamma_{ij}\pi^j} \ , \quad \gamma_{ij}=\partial_i X^I \ 
\partial_j X_I \ 
\end{equation}
and split $N^i$ as 
\begin{equation}
N^i=N_{\bot}^i+\pi^i N_{II} \ , \quad N_{\bot}^i \gamma_{ij}\pi^j=0 \ .  
\end{equation}
If we demand that $\pi^i=\tau_p\tpi^i$ where $\tp^i$ is constant vector 
we again find that 
\begin{equation}
\partial_\sigma N^i_{\bot}=0 \ . 
\end{equation}
We further demand that all world-volume field depend on 
$\sigma$ where $\sigma$ is defined by the relation 
$\partial_\sigma (\dots)=\tpi^i\partial_i(\dots)$. In other
words we require  that $\partial_i (\dots)\triangle^i_{ \ j}=0$. 
Now we are ready to propose following ansatz 
\begin{eqnarray}\label{ansstring}
& &P_\alpha(\xi^0,\sigma)|_{\xi^0=\tau}=\frac{\tau_p}{\tau_F}
K_\alpha(\tau,\sigma)  \ , \quad X^\alpha(\xi^0,\xi^1)|_{\xi^0=\tau}=Z^\alpha(\tau,\sigma) \ , \quad 
\alpha=0,1 \ , 
\nonumber \\
& &P_m=0 \ , \quad Z^m=Z^i(\tau) \ , \quad  m=2,\dots,p  \ ,  \nonumber \\
& &X^I(\xi^0,\sigma)|_{\xi^0=\tau}=Z^I(\tau,\sigma) \ , \quad 
P_I(\xi^0,\sigma)|_{\xi^0=\tau}=\frac{\tau_p}{\tau_F}
K_I(\tau,\sigma) \ ,  \quad 
I=p+1,\dots,D-1 \ ,  \nonumber \\
& & N(\xi^0,\sigma)|_{\xi^0=\tau}=\frac{\tau^2_F}{\tau^2_p}
\lambda_\tau(\tau,\sigma) \ , \quad 
 N^{II}(\xi^0,\sigma)|_{\xi^0=\tau}=\lambda_\sigma 
(\tau,\sigma) \ .  \nonumber \\
\end{eqnarray}
Inserting the ansatz (\ref{ansstring}) 
into equations of motion  
 (\ref{nonBPS})  we obtain 
\begin{eqnarray}
& &\partial_\tau Z^\alpha-2\lambda_\tau\eta^{\alpha\beta}K_\beta-\lambda_\sigma\partial_\sigma X^\alpha=0 \ , 
\quad 
-\partial_\tau K_\alpha+\partial_\sigma (\lambda_\sigma K_\alpha)=0 \ , \quad  \alpha,\beta=0,1 \ ,  \nonumber \\
& &\partial_\tau Z^m=0 \ , \quad K_m=0 \ , \nonumber \\ 
& &
\partial_\tau Z^I-\lambda_\sigma \partial_\sigma Z^I=0 \ , \quad 
-\partial_\tau K_I+2\tau_F^2\partial_\sigma [\lambda_\tau \delta_{IJ}\partial_\sigma 
Z^J]+\partial_\sigma (\lambda_\sigma K_I)=0 \ , \nonumber \\
& & K_\alpha \eta^{\alpha\beta}K_\beta + K_IK^I
+\tau_F^2 (\partial_\sigma Z^I\partial_\sigma Z_I)=0 \ , \quad K_\alpha \partial_\sigma Z^\beta+K_I\partial_\sigma Z^I=0 \  , 
\nonumber \\
\end{eqnarray}
where in the last step we performed projection of the diffeomorphism constraint $\mH_i$ to the direction spanned by $\tpi^i$ and used an anti symmetry of $F_{ij}$ and also used the fact that projection of these constraints to the directions transverse to $\tpi^i$ vanish due to the presumption that all fields depend on $\sigma$ only. In other words (\ref{ansstring})  solve the 
equations of motion (\ref{nonBPS}) on condition that it solves the Carroll string equations of motion which shows that Carroll non-BPS Dp-brane at its tachyon vacuum state is equivalent to the gas of non-interacting Carroll strings with momenta $K_m=0$. Of course, the dynamics of these strings is again trivial since the transverse coordinates do not depend on conjugate momenta and are constant on-shell in the gauge $\lambda_\sigma=0$. On the other hand the physical meaning of the solutions with $K_m\neq 0$ is unclear and deserves further study.

\acknowledgments{This  work  was
	supported by the Grant Agency of the Czech Republic under the grant
	P201/12/G028. }


\end{document}